# EUROPEAN LONGITUDE PRIZES. IV. THOMAS AXE'S IMPOSSIBLE TERMS

**Richard de Grijs**
*Department of Physics and Astronomy, Macquarie University,
Balaclava Road, Sydney, NSW 2109, Australia*
Email: richard.de-grijs@mq.edu.au

**Abstract:** Although governments across Europe had realised the need to incentivise the development of practically viable longitude solutions as early as the late-sixteenth century, the English government was late to the party. An sense of urgency among the scientific community and maritime navigators led to the establishment of a number of longitude awards by private donors. The first private British award was bequeathed in 1691 by Thomas Axe, parish clerk of Ottery St. Mary (Devon). Despite the absence of an expenses component and the onerous and costly nature of its terms and conditions, the Axe prize attracted a number of optimistic claimants. Although the award was never disbursed, it may have contributed to the instigation of the government-supported monetary reward associated with the British Longitude Act of 1714. It is likely that the conditions governing the British 'Longitude Prize', specifically the required accuracy and the need for sea trials and of disclosure of a successful method's theoretical principles, can be traced back at least in part to the Axe Prize requirements.

**Keywords:** longitude determination, British longitude awards, Thomas Axe, British Longitude Act

## 1 PRIVATE DONORS

Although the British Longitude Act and its associated monetary award of 1714 are well-known and well-documented (e.g., Dunn and Higgitt, 2014; Mobbs and Unwin, 2016), to the extent that the 'Longitude Prize' has captured the public's imagination (e.g., Sobel, 1995), the quest for a viable means of determining one's longitude at sea precedes the announcement of the British award by several centuries. In a series of recent papers, I have reviewed a number of the European government-supported longitude awards which were in place long before the British Parliament passed its 1714 Act. These included the Spanish and Dutch rewards of the late sixteenth and early seventeenth centuries (de Grijs, 2020a, 2021a; Papers I and II), as well as an alleged prize ostensibly offered by the Venetian Republic (de Grijs, 2021b; Paper III).

However, efforts to develop a means of position determination at sea had proceeded apace across the European maritime nations well before the first awards were formally announced. Private individuals joined the fray by announcing their own longitude awards (e.g., Gould, 1923: 12; Hall, 1930). Perhaps the best contemporary example of such a private donor is the case of Count Jean-Baptiste Rouillé de Meslay (1656–1715), who in March 1714 endowed a prize of 125,000 *livres* (equivalent to £6,250 in 1714 pounds sterling; Howse, 1997: 61). The Meslay prize was to be awarded by the Paris *Académie des sciences*, in part to incentivise developments in longitude determination (e.g., Boistel, 2015; Bret, 2019).

Despite the traditionally intense Anglo–French rivalry, the Meslay prize appears to have been established without direct reference to the contemporaneous British prize on offer. Both Britain and France were actively trying to solve the era's great navigational problems. However, while Britain's focus was on solving the longitude problem, France's quest was the Earth's shape (e.g., Boistel, 2015). De Meslay bequeathed prize funds to the *Académie des sciences* for two philosophical dissertations; the smaller of these prizes was meant for the person

> … who best achieved the shortest and easiest method and rule for taking the heights and degrees of longitude at sea exactly and [for] useful discoveries for navigation and great voyages. (Maindron, 1881: 15; transl.: Howse, 1997: 61).

Although somewhat hidden in the bequest's description above, the smaller Meslay prize was not awarded exclusively for longitude-related achievements.

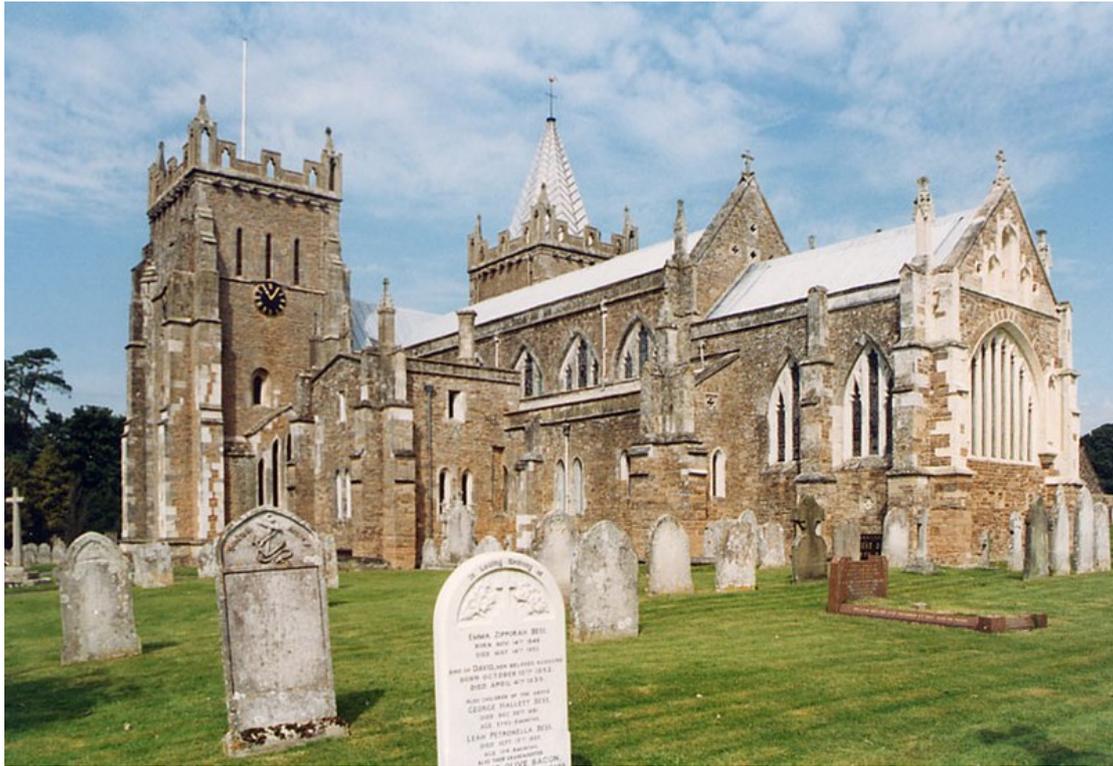
**Figure 1.** St Mary's Church from the South East. Ottery St. Mary, Devon. (John Salmon; CC BY-SA 2.0)

The first privately endowed scientific prize in France did not see the light of day until 1712, when Henri-Jacques Nompar de Caumont (1675–1726), Duc de La Force, worked with a provincial academy in Bordeaux to endow a prize. It was first bestowed in 1715 (Bret, 2019). Shortly afterwards, in 1716 Philippe Charles II (1674–1723), Duke of Orléans and Regent of France, promoted his own private longitude award, although for an unspecified amount (Howse, 1997: 62). The latter prize was never awarded.

Two decades earlier, however, Thomas Axe (1635–1691), parish clerk of Ottery St. Mary's church (see Figure 1) in Devon, had already bequeathed £1,000 in his will—dated 20 July 1691, "… written on five sheets of paper …" (Hall, 1930)—to reward the development of a longitude method whereby men "… of mean capacity …", that is, not among society's brightest (see below), could determine longitude at sea to an accuracy of better than half a degree (formally as measured at the Equator).

Although the Axe Prize is not well-known outside of niche scholarly circles, some scholars contend that it contributed, perhaps significantly, to the developments leading up to the announcement of the British Longitude Award of 1714. On the other hand, it is clear that Axe knew many of the leading scientists of the day and he was, therefore, likely involved in ongoing discussions as to how an incentive prize might be formulated. Those discussions could very well have been reflected in his will.

Axe's endowment preceded the British Longitude Award by almost a quarter of a century, at a time when developments and new proposals to solve the perennial longitude problem mushroomed and caught the public's imagination. In the late-seventeenth century, King Charles II (reigned 1660–1685) received increasing numbers of proposals from numerous scholars and opportunists claiming to have solved the problem of finding the 'true longitude'. Upon having received proposals from a French explorer known as Le Sieur de Saint-Pierre, most likely Jean-Paul Le Gardeur (1661–1738; de Grijs, 2017: Ch. 6), and from a certain 'old' Henry Bond (ca. 1600–1678), author of *The Longitude Found* (1676), on 15 December 1674 Charles II appointed a Royal 'Longitude Commission' to scrutinise these proposals:

> … that he [St. Pierre] hath found out the true knowledge of the Longitude, and desires to be put on Tryall thereof; Wee having taken the Same into Our consideration, and being willing to give all fitting encouragement to an Undertaking soe beneficiall to the Publick … hereby doe constitute and appoint you [the Commissioners], or any four of you, to meet together …
>
> And You are to call to your assistance such Persons, as You shall think fit: And Our pleasure is that when you have had sufficient Tryalls of his Skill in this matter of finding out the true Longitude from such observations, as You shall have made and given him, that you make Report thereof together with your opinions there-upon, how farre it may be Practicable and usefull to the Publick. (British Library, 1674).

Although strongly criticised by Robert Hooke (1635–1703; de Grijs, 2017: Ch. 6), Bond's proposal based on magnetic declination (Taylor, 1939) resulted in the only Royal award ever approved (Shaw, 1676–1679), a life annuity of £50.

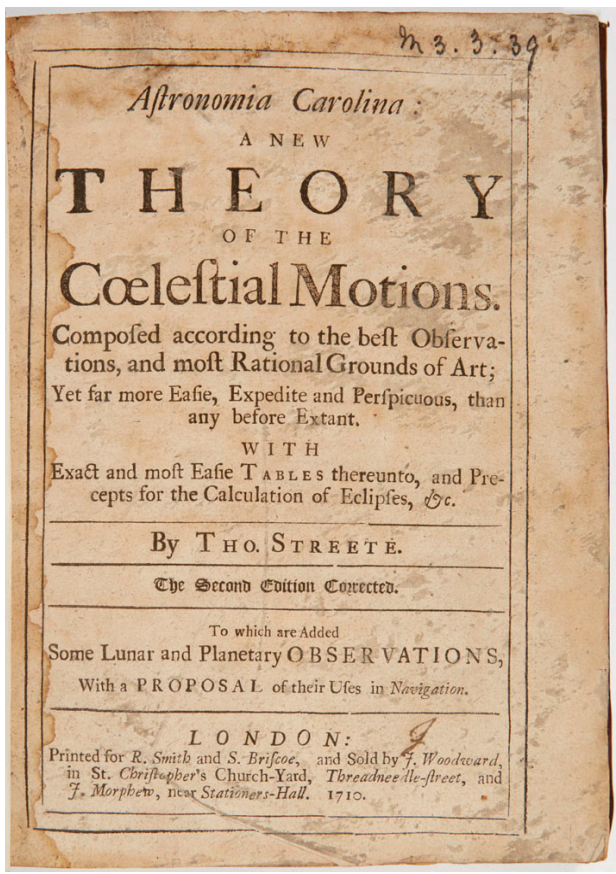

**Figure 2.** Thomas Streete's 1664 *Astronomia Carolina*; second edition (1710). (Wellcome Trust; Creative Commons Attribution 4.0 International license)

This late seventeenth-century round of frenetic activity to find a solution to the longitude problem eventually resulted, in part, in the foundation of the Royal Greenwich Observatory in 1675. In turn, this triggered the publication of increasing numbers of pamphlets containing longitude 'solutions', either real or imagined. The ensuing public frenzy led to Axe's endowment in 1691.

An important reason for private donors to enter the longitude game may have been the general absence of Government incentives (that is, monetary rewards) to develop a practically viable solution (e.g., Turner, 1996; Wepster, 2000: 8–9, 14). Indeed, it took until the second half of the seventeenth century before English transoceanic commerce became sufficiently economically important for the country's politicians to take note of the thorny longitude problem (e.g., Turner, 1996; and references therein). Contrary to the political environment in France (Pares, 1980), there was no sense of urgency in seventeenth-century Britain, nor any concerted effort, to find a practical solution to the problem.

Given this lackluster political response, some inventors were loath to disclose their solutions in the absence of generous incentives for their efforts. As a case in point, as early as 1664, the English scholar Thomas Streete (1622–1689) announced that he had developed a longitude solution (Streete, 1664: see Figure 2). Streete's approach was, in essence, a form of the lunar distance method (see de Grijs, 2020b) using the Moon's zodiacal position. However, he was unwilling to publish the details unless a monetary reward for his invention's continued development would be forthcoming (e.g., Turner, 1996; Wepster, 2000).

**2 THOMAS AXE, THE MAN**

Thomas Axe was born in December 1635. He was the eldest son of George Axe (Cornish, 1869: 46; but see Turner, 1996: notes 32, 33). Nothing is known about his early life, although he must have enjoyed a solid education in mathematics from a young age, likely under the tutelage of his father (Otis, 2013: 121). Axe was taught "… the table of Multiplication when he was seven years old." (Aubrey, 1669–1684: f. 29r). This stood him in good stead, since (Otis, 2013: 122; citing Aubrey, 1669–1684: f. 29r)

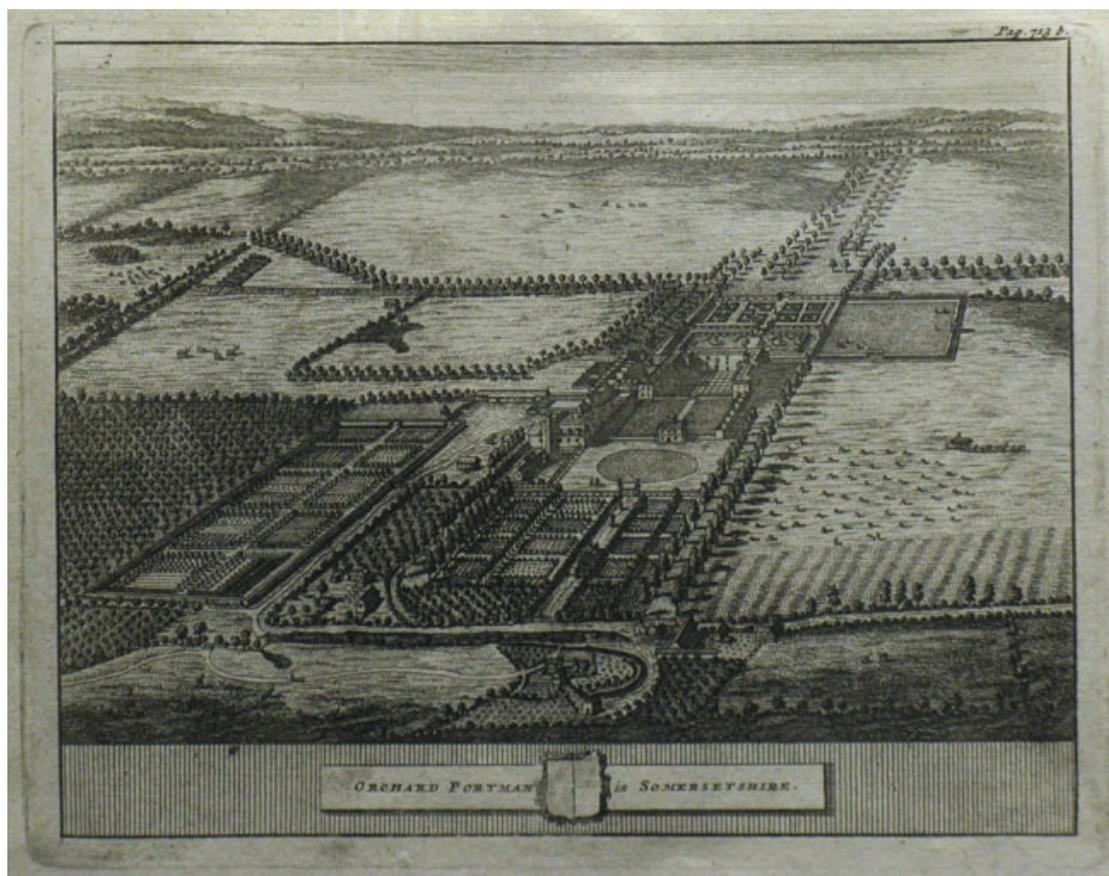

**Figure 3.** Orchard Portman, painted by Pieter van der Aa about 1707. (Wikimedia Commons; public domain)

> Boies will Adde, Mutiply & Diuide as
> fast as a Dog will trot: will run up an account
> Like a shop-keeper. A Barre-boy at an
> Ale-house will reckon faster & readier than
> a Master of Arts in a University, or a Justice of Peace.

He grew up as a 'Gentleman' of relative wealth in Orchard—also known as Orchard Portman—in the county of Somerset, near Taunton (see Figure 3). When he reached the age of 34, he married Dorothy Bristall, daughter of the Master of the King's School, who had formerly been Master of the school at Blandford, Dorset. The register of marriage licences issued by the Faculty Office of the Archbishop of Canterbury (Chester, 1886: 114) shows,

| Aug. 24 [1670] | Thomas Axe, of Orchard, co[unty] Somerset, Gent[leman], Bach[elor], 34, & Dorothy Birstall, Sp[inster], about 24, dau[ghter] of William Birstall, D.D. [Doctor of Divinity], of Eversley, co[unty] South$^{ton}$ [Southampton], who consents; at Eversley afs$^d$ [aforesaid]. |

Axe was appointed assistant estate steward, under one Mr. Colby, to administer the lands of Sir William Portman (1643–1690), 6$^{th}$ baronet, of Bryantstone, near Blandford, probably through the interests of his father-in-law (Cornish, 1869: 46). Upon Colby's death, Axe became the sole steward of the property and subsequently moved to Orchard Portman. Given his keen interests in finances and statistics, he went on to also provide investment advice as attorney to Sir William (Axe, 1684). The baronet, meanwhile, served as Axe's local member of Parliament for Somerset between 1679 and 1685 (London Public Record Office, 1690). Upon the baronet's death, Axe was asked by the English natural philosopher and educational theorist John Aubrey (1626–1697) to provide his 'Calculations of Sir William Portman's Old Rents', which he had used as the basis for his investment advice (Stephens,

1972: 104; Bennett, 2009: 341).

Through his lifelong interest in statistical applications, and arithmetic in general (for details, see Turner, 1996), he became acquainted and made friends with members (Fellows) of the newly established Royal Society of London, including Sir John Cutler (1603–1693), John Graunt (1620–1674), Sir William Petty (1623–1687), Aubrey and Hooke. And so we can be certain that he was well aware of the latest scientific ideas and developments coming out of London—including the ever-increasing interest of the scientific establishment in the longitude problem. In turn, this likely prompted him to endow a longitude prize.

## 3 THOMAS AXE'S WILL

Here I review the terms of the Axe Prize and the context in which these must be considered. My aim is to provide a comprehensive overview of developments associated with the Axe Prize, given that the literature is widely scattered and often hard to find, whereas a few key references are still only available in hardcopy (e.g., Howse, 1980, 1997; Turner, 1996).

Following Axe's passing, the first mention of his longitude legacy is found in a letter of 13 (19?) September 1691 from Richard Lapthorne (fl. 1687–1697) to Richard Coffin (d. 1700) in Portledge, Devon:

> Mr. Axe, Sir William Portman's servant, whom I knew well, lately dyed and by his will hath given 1000 *li*. to incourage some mathematicall students to finde out the true longitude, for the benefit of navigators. (Historical Manuscripts Commission, 1876: 382a; Kerr and Coffin Duncan, 1928: 121)

Next, in 1696 Edward Harrison (fl. 1686–1700), a lieutenant in the British Royal Navy, briefly referred to the Axe prize in his treatise, *Idea longitudinis, being a brief definition of the best known axioms for finding the longitude*. Harrison (1696: 76–77) was clearly not impressed. In the excerpt below, he levelled scathing criticism at the Axe Prize (emphasis mine), which was matched by equally scathing reviews of the (alleged) Venetian and Dutch prizes (see Paper III):

> As I have heard say, … Thomas Axe, an Englishman, left a Legacy of One Thousand Pound, (**never to be paid I think**,) to any Person that should discover the Longitude, within the space of Ten Years after his Decease, if his Wife and Child died Childless in that time; besides, it is to be approved of by the four Professors of Geometry and Astronomy, in Oxford and Cambridge, for the time being, and at least twenty Experienc'd Masters of Ships, that shall have made several Experiments thereof in long Voyages. Affidavits are to be made before the Twelve Judges of England, &c. He dyed in the year 1691, and I think took care enough, that the said one Thousand Pound should be Irrecoverable; … (Harrison, 1696: 76–77)

In order to understand the reason for Harrison's criticism, I include the relevant text of Axe's will below in full to provide context:

> Whereas I am convinced by several of great knowledge in Astronomy and Cosmography, that they know how to make the Masters of Ships of the meanest capacity [not among the brightest] to find out the *Longitude* as readily, and almost exact as they can find the *Latitude*, but do not communicate this useful secret to the World, seeing it hath and will cost them much money and pains by making instruments, and therefore do expect a Reward from the Publick, or others, for the Discovery:–

> Wherefore, if any person or persons shall assist within *Ten years* next after my decease as aforesaid, make such perfect discovery how men of mean capacity may find out the Longitude at Sea, so as at any time they can truly pronounce upon observation as within *half a Degree* of the true Longitude, and shall demonstrate how it may be done to the Two Professors of Geometry in Oxford and Cambridge for the time being, and shall give ample proof thereof by the Affidavits in writing of at least Twenty able Masters of Ships, that shall have made several experiments thereof in long Voyages, which Affidavits are to be made before the Twelve Judges of England or before the major part of them, of which the Lords Chief Justices to be Two, and shall deliver a Certificate thereof signed and sealed by the said major part of the said Justices, unto the said [Trustees] Samuel Keeble [bookseller of St. Dunstan's parish, West London], William Pratt, and Henry Hooper, and the Survivors of them and the Heirs of such

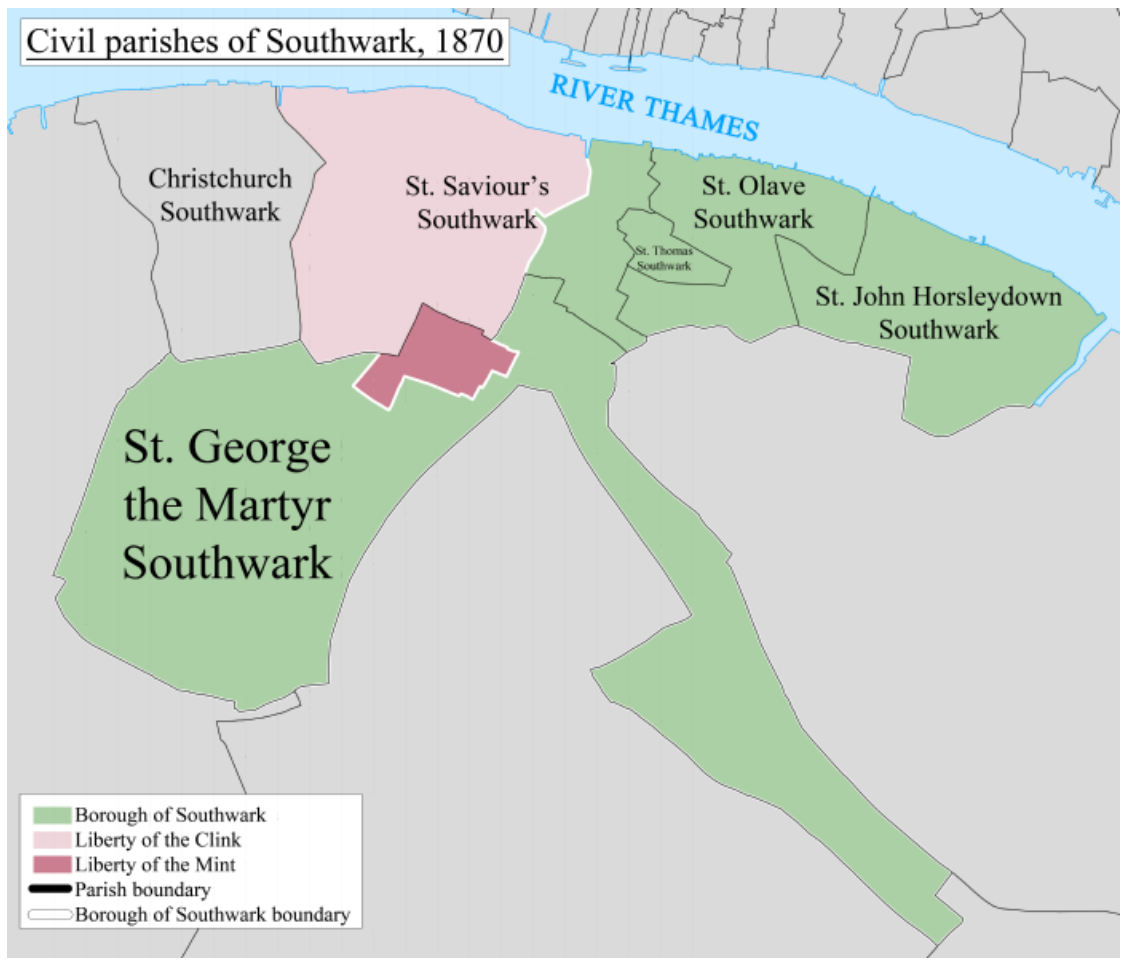

**Figure 4.** Civil parishes of Southwark (1870). The 'Clink' in the figure's legend is the historic Southwark prison (1151–1780); the Liberty of the Clink was a local manor area owned by the Bishop of Winchester rather than by the reigning monarch. (Doc77can via Wikimedia Commons; Creative Commons Attribution-Share Alike 3.0 Unported license)

Survivors, with all convenient speed next after the death of me, my Wife [Dorothy], and Son [Henry] childless, sell the Fee of all my said Houses in the said Parish of St. Olave and St. Saviour's [Southwark, London; see Figures 4 and 5], for as much money as can be gotten for the same, and that the sum of One Thousand Pounds of such purchase money be paid unto such person or persons as shall have so discovered such ready and easy way of finding *the true* Longitude at Sea, and shall have made such proof as aforesaid for the benefit of Mankind, which said sum of One Thousand Pounds, on the conditions and in the case aforesaid, I give and bequeath to such person or persons accordingly;–

And moreover my Will is, that from and immediately after the death of my said Wife, and of myself, and said Son leaving no issue [offspring] of either of our bodies behind us as aforesaid, that they Samuel Keeble, William Pratt, and Henry Hooper, and the Survivors or Survivors of them the Heirs of such Survivors, shall once in every year, until perfect proof of the Discovery of the true Longitude shall be as aforesaid, and the said Houses remain unsold, pay Forty Pounds, part of the rents of the said Houses, unto and amongst such person and persons for their encouragement, as shall have employed their time to make Observations at several Islands, and noted places in Europe, Asia, Africa, and America, according to the Rules and just (description) of such said person or persons shall have made such discovery, how all men may find out the Longitude, in order of making of an accurate map of the World as to *Longitude* and *Latitude*,–the disposal of the sum of £1000., and of the said £40. by the year, my good friend [Astronomer Royal] Mr. Edmund Halley, and [Professor of Geometry at Gresham College] Mr. Robert Hooke, be consulted.–

And my Will is, that if such perfect discovery of the Longitude as aforesaid, *shall not be made* within *Ten years* after the death of me, my Wife, and our said Son childless, that in such case the said Houses in Southwark *shall not be sold*, but that the clear yearly profits of the same shall be divided into Twelve equal parts, and be distributed near about *St. Thomas's* day

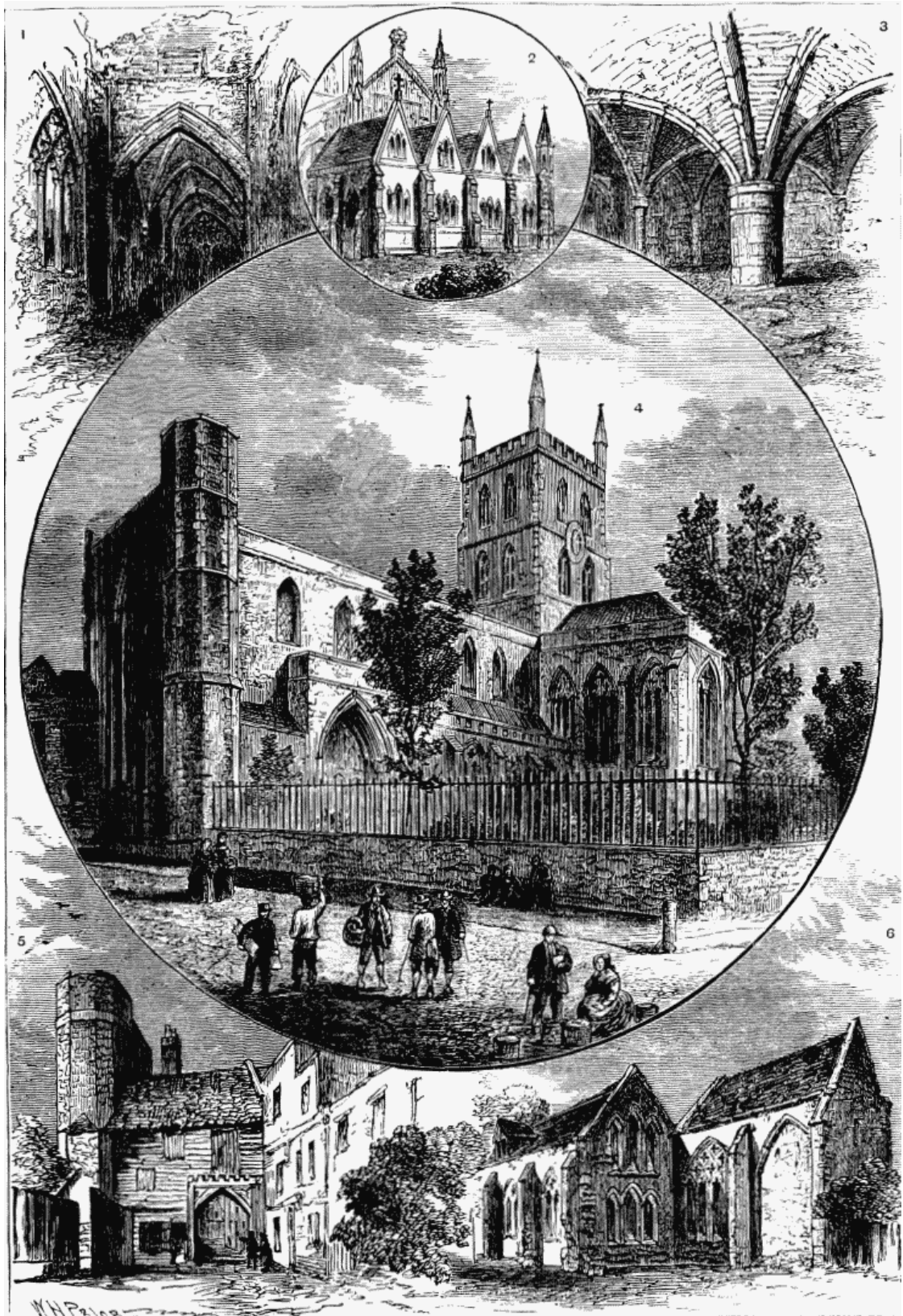

**Figure 5.** Views of St. Saviour's church. 1: Chapel interior, East End of St. Saviour's; 2: Lady Chapel; 3: Part of St. Saviour's Priory; 4: St. Saviour's Church; 5: Montague Close; 6: Chapel at the end of St. Saviour's. (via British History Online; CC BY NC SA 2.0)

[Winter solstice, 21 December] to and amongst my said Brothers and Sisters and Kindred during their respective lives, in such portion and portions as I have limited my personal estate to them respectively in and by this my Will,–And from and after the death of my said Brothers and Sisters and Kindred, I give the said last mentioned Houses to The Governors of the goods and chattels of The Collegiate Church of Ottery St. Mary aforesaid, Upon Trust, that they will yearly about *St. Thomas's* day distribute the same, as followeth,–*viz*. One part of the last mentioned Houses [to the Vicar's wife], to buy drugs and plaisters [for the poor],–One part to the Minister,–One part to the Vicar,–One part to the Grammar School-Master,–One part to the Parish Clerk of Ottery St. Mary [i.e., Axe's successor],–And as a Fund to encourage Steadiness, Sobriety, and Industry, *Three* of those parts to be as a Stock, out of which, upon the Marriage day of any Young Man or Woman that shall not marry 'till after the age of One and Twenty [twenty-one] years, and that shall have lived in one Service in the Parish of Ottery St. Mary for *seven* years, and shall not have had Alms of the Parish or Ottery St. Mary within *five* years next before such Marriage, the sum of £3. &c.– (Carlisle, 1818: 328–330; Hall, 1930)

Although the will's first paragraph above implies that Axe seems to have been convinced that the longitude problem was all but solved, the 'Axe Prize' was never paid out. In fact, the implication of the language in Axe's will is that the scientific establishment was really after a means to make one of the proposed, existing methods of longitude determination at sea work in practice, rather than pursue the invention of a completely new, as yet untested method. Separately, it is unknown whether the annual incentives (to be sourced from the rent of his properties) of up to £40 for scholars involved in making corrected world maps were ever paid out.

## 4 TERMS AND CONDITIONS

It is likely that Axe's reward was "… coupled with such an absurd number of stipulations as made it, whether intentionally or otherwise, practically impossible to win." as eloquently elaborated upon by Gould (1931; see also Gould, 1923: 12–13):

> ... the reward can never be paid at all unless Axe, his wife and his son all die childless—a condition impossible of fulfilment [*sic*] unless Axe's son (or sons—the will is ambiguous on this point) predeceases both his father and his mother. If it were an ordinary matter of succession, no sane man would expect to get £10 for his expectation of succeeding to a reversion of £1,000 under such conditions.

> But, assuming this difficulty surmounted, a claimant's troubles are only beginning. He must send his invention for several long voyages with each of twenty masters of vessels; a proceeding which, if he can only afford one sample of the invention, may take some sixty years or so [this seems somewhat exaggerated]. He must then obtain—and presumably pay for—twenty affidavits from those masters, all of which affidavits must be favourable. Armed with these, if he has any money and patience left, he must next repair to Oxford and Cambridge and, by some means or other, convince four professors (who know probably nothing of his *bona fides*; and, possibly, very little of navigation) that his invention does what it claims to do—a feat which, as he has confessed, baffled even John Harrison [the Yorkshire clockmaker who was eventually recognised as the person who solved the longitude problem]. Next, our claimant must swear, and pay for, a lot more affidavits before "the Twelve Judges of England," or the majority of them; and from them he must obtain a certificate (which, as a matter of judicial practice, I imagine that they would be exceedingly loth to grant) addressed to Axe's executors, if any of them are still alive. He may then possibly receive his £1,000, having spent, in quest of it, a considerably larger sum—always provided that some relative of Axe does not object and throw the whole business into Chancery [the court of equity].

Indeed, John Harrison spent more than three times as much as the potential Axe reward—£3,250—in preparation of the submission of his timepieces to the Commissioners of the Longitude. At various times throughout his long and arduous journey to eventually construct a maritime clock that satisfied the terms of the British Longitude Act, he received advances from the Board of Longitude to help him move forward. Similarly, the Board of Longitude paid out advances and expenses to a number of applicants whose inventions showed promise, including to Leonhard Euler in May 1765, to George Witchell in July 1765, to Israel Lyons in October 1765, to Richard Dunthorne in 1771, to Tobias Mayer's widow, also in 1771, to Charles Mason in 1787 and to William Lax in 1828 (for a full list of awardees see, e.g., Howse, 1998: Appendix; see also de Grijs, 2020b).

As such, I suspect that whereas Axe may have been up to date with contemporary scientific developments, he was likely too far removed from the day-to-day reality of practical navigation to understand what was involved. That may have led to the overly onerous and costly ('impossible') terms and conditions associated with his otherwise benevolent bequest.

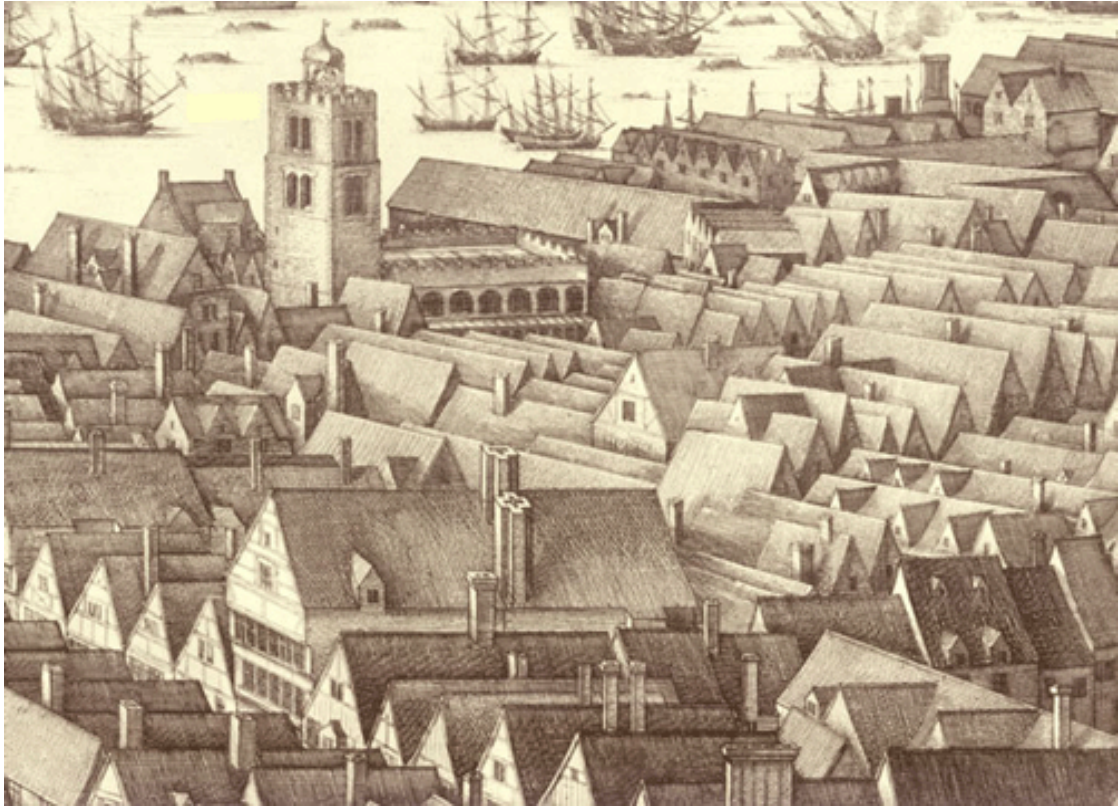

**Figure 6.** St. Olave's Church, Tooley Street. The river Thames is visible at the top of this 1647 drawing by Wenceslaus Hollar. (public domain)

## 5 ENACTMENT

Thomas Axe died in 1691 and was buried at Orchard Portman on 1 August of that year (Cornish, 1869: 46),

> And I direct my Body to be buried in some Churchyard within Ten Miles of the place where I shall happen to die but not until seven days next after my decease with as little pomp and noise as may be. (Hall, 1930: 66)

Although the headstone of his grave remained unmarked, a lengthy inscription recording the conditions of Axe's longitude award appears on a 'debased Gothic' cenotaph covering the entire western wall of the south porch of the Church of Ottery St. Mary (Hall, 1930). His family appears to have passed away fairly soon afterwards. This suggestion is based on the notion that Edward Harrison (1696: 77–78) complained that, despite having constructed unspecified 'instruments' in an attempt at obtaining the Axe prize, the £1,000 prize money was irrecoverable:

> I know not how to recover my Cost and Charges, though I have laid (and shall the rest by and by,) most of the ground work at my Charge; where is the Money to raise and perfect the Work? There is — Pounds wanting, present Pay, towards the Charge of making two wonderful Instruments, for the first making of the Instrument, will be very Chargeable; one of them may prove far more chargeable then the other, and yet when it becomes common, it may be sold at a low Price, I Judge under fifty Shillings; yet before the *Instrument* can be brought to such Perfection, as to show two or three *Examples,* may cost several Hundreds of Pounds, and for ought I know, take up most part of a Year, to accomplish them, it is an *Instrument* that must be truly Cut and Polished, a small matter of a false sweep or stroke may quite spoil it, and make it good for little or nothing; therefore one Hundred of them may happen to be spoiled before we can get the right Art of making them; neither of them will suddenly be fit for Seamen's Practise, because the *Natural motions* that one of them is to show, are at this present writing, I believe, unknown to the World; therefore reasonable time is required to Calculate the *Theories* of those *motions.*

Axe had stipulated in his will that, upon his death, his houses in Southwark and elsewhere in Surrey would be passed to the care of his Trustees ("four Governors", which eventually became the Church Corporation Trust) and their heirs, for the benefit of his wife during her lifetime, next—upon his wife's passing—of their son during his lifetime, and then down to their heirs. In case no children remained after the deaths of both his wife and son, the Trustees were to sell those houses and use the proceeds to endow a longitude prize. If a longitude solution would not be forthcoming within ten years of Axe's death, he stipulated that the rental proceeds from his houses should be divided into twelve parts and split among a number of beneficiaries, as indicated in Section 3 above.

Axe's houses in Southwark included three properties in Tooley Street (see Figure 6), which were sold on 30 January 1790 for £900 (after costs) to allow for widening and improvements of a number of streets (Commissioners for Inquiry into Charities, 1826: 57; 1839: 108), as well as two houses in the Borough's High Street, which were rented to tenants. The latter included "… the Bell alehouse, numbered 314, to which is attached a messuage [a dwelling house with outbuildings and land], situate[d] in Pepper-alley; and the other a dwelling house numbered 315." (Commissioners for Inquiry into Charities, 1826: 57; 1839: 108).

Back in Orchard, some of Axe's houses were meant to benefit the schoolmaster, although they were sold shortly after Axe's death. The proceeds of their sale were invested in a fund managed by his Trustees, who paid the Master about £8 8*s*. from the dividends (Carlisle, 1818: 331; Randell, 1999), with another £8 8*s*. each going to the vicar, chaplain and parish clerk, and the same amount spent on drugs for the poor (Randell, 1999). By 1822, the "clear income" from Axe's legacy was £100 18*s*. 4*d*. per annum (Lysons, 1822: 380). In addition, Axe bequeathed the rental proceeds of an estate in Blandford, Dorsetshire, comprising a house, three cottages and land, with three quarters allocated to the parish clerk and the remainder meant to provide medical or surgical assistance to the poor (Commissioners for Inquiry into Charities, 1826: 57; 1839: 108). By 1822, these returns were paid to the Exeter and Devon Hospital (Lysons, 1822: 380).

By 1850, the managed fund's dividends had increased to £1,426 5*s*. 10*d*. (Randell, 1999). The Thomas Axe non-ecclesiastical charity remains active today. It aims at poverty prevention and relief for the elderly in the parish of Ottery St. Mary. Its three Trustees still represent the Ottery Church Corporation, alongside a small number of other entities. The charity's income reported for the financial year ending on 31 December 2019 was £1,769 (Charity Commission for England and Wales, 2021).

**6 BRITISH LONGITUDE CONTEXT**

Despite the practically impossible terms of the Axe reward (Gould, 923: 12–13; Gould, 1931), its basic framework comprising three award levels based on the eventual accuracy of any proposed and viable method was subsequently adopted by the Cambridge mathematicians William Whiston (1667–1752) and Humphrey Ditton (1675–1715) in their petition to the British Parliament (e.g., de Grijs, 2017: Ch. 6.4.2). This eventually led Sir Isaac Newton (1642–1726/7) and Edmond Halley (1656–1752), among others, to recommend the same to the parliamentary committee. It is at present impossible to confirm whether the Axe Prize's terms and conditions directly impacted on those later proposals, or whether the three award levels were simply the result of applying common sense.

In any case, Halley was a member of the commission established to assess the Whiston–Ditton proposal, and he would also have been familiar with the conditions tied to the Axe award. Moreover, the requirements of the eventual British Longitude Award of 1714 were the same as those imposed by Axe in his will. First, the highest tier of award money would be disbursed to the person who could design a method of longitude determination that was better than half a degree (at the Equator). Second, both Axe's will and the British Longitude Award required sea trials to demonstrate practical viability. Third, both prizes required disclosure of the underlying theoretical principles to competent authorities for independent assessment. Note that the Professors of Geometry at Oxford and Cambridge enlisted by Axe were also *ex officio* Commissioners of the Longitude (e.g., de Grijs, 2017: Table 6.1).

One might be somewhat bemused by Axe's requirement to have the results validated by the chief justices of England, but that requirement simply reflected his experiences and routine approach as a gentleman with certain commercial interests (e.g., Turner, 1996). Despite the excessive requirements imposed by Axe's will, at least some 'projectors' were optimistic that the prize was theirs for the taking:

> I'll tell you of a great discovery. There's one pretends y$^t$ he has found ye long-studied secret of taking the Longitude by sea. He has communicated with Sir Is. Newton & Mr. [John] Flamstead [*sic*], and is preparing a large instrument to make y$^e$ experiment, & will go down to York to try it, it is not impossible y$^t$ he may succeed at sea if what he says is true y$^t$ he can do it wth ease & certainty as he walks on foot by looking at some stars thro his instrument. When he has sufficiently experienced it he designs to go to my L$^d$ Pembrook [Thomas Herbert, 8$^{th}$ Earl of Pembroke and President of the Royal Society of London; ca. 1656–1733] to gett a Proclamacion for a publickk reward, & to appoint him wt Judges y$^e$ Govern$^t$ pleases to try it & publish it to [*sic*]. Besides his expectations from y$^e$ Pub. he is sure of a praemium of 1000 *l*. left in will by a Gent. for that purpose, this account I had from one of y$^e$ Trustees of that will, who offer'd to shew it me. (Smith, 1784)

On the basis of this anonymous account, Turner (1996) has suggested that the executors of Axe's will may have displayed some flexibility in the disbursement of the monetary reward, particularly as regards the timing. After all, Axe's will stipulated that the prize money could only be paid out within ten years of his death, whereas the passage quoted above was taken from a letter sent in 1706. Nevertheless, Axe's requirements proved too arduous for even the most persistent claimant. Yet, despite this setback, his offer of a longitude prize was a sign of the times and of the urgency felt by the London scientific community. The British Longitude Act would soon propel the nation into a major frenzy, although a longitude solution would not be formally affirmed until 1773, when the British Parliament finally granted John Harrison his long-coveted award.

## 7 ACKNOWLEDGEMENTS

I gratefully acknowledge assistance from Keith Piggott in sourcing a few hard-to-find historical articles. I also acknowledge a number of helpful suggestions by the paper's reviewers.